\documentclass[11pt]{article}
\usepackage{hyperref}
\usepackage{scalerel}
\usepackage[normalem]{ulem}
\usepackage{amsfonts}
\usepackage{bm}
\usepackage{cite}
\usepackage{mathrsfs}
\usepackage{amsmath}
\usepackage{graphicx}
\usepackage{hyperref}
\usepackage{feynmp-auto}
\usepackage{verbatim}
\usepackage{color}
\usepackage[dvipsnames]{xcolor}
\usepackage{enumerate}
\usepackage{amsfonts}
\usepackage{amssymb}
\usepackage{authblk}
\usepackage[hang,flushmargin]{footmisc}
\usepackage{lipsum}
\usepackage[font=footnotesize]{caption}
\usepackage{soul}
\DeclareMathAlphabet\mathbfcal{OMS}{cmsy}{b}{n}
\usepackage{empheq}
\usepackage{cancel}

\csname @addtoreset\endcsname{equation}{section}

 \newcommand{\badat}{\begin{alignedat}}
 \newcommand{\eadat}{\end{alignedat}}

\def\be{\begin{eqnarray}}
\def\ee{\end{eqnarray}}
\def\beann{\begin{eqnarray*}}
\def\eeann{\end{eqnarray*}}
\def\beq{\begin{equation}}
\def\eeq{\end{equation}}
\def\ba{\begin{array}}
\def\ea{\end{array}}
\def\ben{\begin{enumerate}}
\def\een{\end{enumerate}}
\def\bea{\begin{align}}
\def\eea{\end{align}}

\def\5{\bar }
\def\6{\partial }
\def\7{\hat }
\def\4{\tilde }

\renewcommand{\d}{\partial}

\def\cD{\mathcal{D}}

\def\cF{\mathcal{F}}
\def\cG{\mathcal{G}}
\def\cH{\mathcal{H}}

\def\cL{\mathcal{L}}

\def\cN{\mathcal{N}}

\def\cR{\mathcal{R}}
\def\cS{\mathcal{S}}

\definecolor{blue-violet}{rgb}{0.54, 0.17, 0.89}
\definecolor{PineGreen}{cmyk}{0.92, 0, 0.59, 0.25}
\definecolor{OliveGreen}{cmyk}{0.64, 0, 0.95, 0.40}
\definecolor{RawSienna}{cmyk}{0, 0.72, 1, 0.45}
\definecolor{Gray}{cmyk}{0, 0, 0, 0.50}
\definecolor{MidnightBlue}{cmyk}{0.98, 0.13, 0, 0.43}
\definecolor{Orange}{cmyk}{0, 0.61, 0.87, 0}
\definecolor{LimeGreen}{cmyk}{0.50, 0, 1, 0}
\definecolor{Green}{cmyk}{1, 0, 1, 0}

\renewcommand{\d}{\partial}
\renewcommand{\tilde}{\widetilde}
\renewcommand{\hat}{\widehat}
\def\beq{\begin{eqnarray}}
\def\eeq{\end{eqnarray}}
\def\beann{\begin{eqnarray*}}
\def\eeann{\end{eqnarray*}}
\def\be{\begin{equation}}
\def\ee{\end{equation}}
\def\ba{\begin{array}}
\def\ea{\end{array}}
\def\ben{\begin{enumerate}}
\def\een{\end{enumerate}}
\def\bea{\begin{eqnarray}}
\def\eea{\end{eqnarray}}

\def\dD{\mathscr{D}}

\def\dI{\mathscr{I}}

\hypersetup{final}

\begin{document}
\begin{titlepage}

\vspace{30pt}

\begin{center}


{\Large\sc Scalar subleading soft theorems from \\[10pt]   an infinite tower of charges}

\vspace{-5pt}
\par\noindent\rule{350pt}{0.4pt}


\vspace{20pt}
{\sc 
Matías Briceño${}^{\; a,}$,
Hernán A. González${}^{\; a}$ and
Alfredo~P\'erez${}^{\; b,c}$
}

\vspace{8pt}

${}^a${\it\small 
Facultad de Ingeniería, Universidad San Sebastián, Santiago 8420524, Chile.}
\vspace{4pt}

${}^b${\it\small
Centro de Estudios Cient\'ificos (CECs), Avenida Arturo Prat 514, Valdivia, Chile.}
\vspace{4pt}

${}^c${\it\small
Facultad de Ingenier\'ia, Universidad San Sebasti\'an, sede Valdivia, General Lagos 1163, Valdivia 5110693, Chile.}
\vspace{4pt}

{\tt\small 
\href{mbricenoc3@correo.uss.cl}{mbricenoc3@correo.uss.cl},
\href{mailto:hernan.gonzalez@uss.cl}{hernan.gonzalez@uss.cl},
\href{mailto:alfredo.perez@uss.cl}{alfredo.perez@uss.cl}
}
\end{center}
\begin{abstract}
\noindent We investigate conservation laws in the absence of gauge invariance by analyzing interacting scalar theories with a massless scalar field. We construct an infinite tower of charges that arise from the subleading equations of motion at null infinity and are built from specific combinations of asymptotic field coefficients. Interestingly, these expressions are finite from the outset, requiring no holographic renormalization.
By carefully analyzing the dynamics at spatial infinity, we show that this tower of surface integrals commutes with the $S$-matrix of the interacting model at tree level. As an application, we demonstrate that the conservation of these charges leads to an infinite set of subleading soft relations, valid at tree level in a cubic interaction with massive scalar fields. 
\end{abstract}

 \end{titlepage}
\tableofcontents

\section{Introduction}
Soft theorems\cite{Low:1954kd,Low:1958sn,Weinberg:1964ew,Weinberg:1964kqu,Weinberg:1965nx} are manifestations of the universal behavior of long-range interactions. Over the past years, they have evolved into a rich framework for exploring symmetry enhancements in the asymptotic region of spacetime, unveiling connections with the holographic principle in flat spacetime \cite{deBoer:2003vf,Cheung:2016iub,Strominger:2021mtt,Guevara:2021abz} and the underlying infrared structure of gravitational and gauge theories (see \cite{Strominger:2017zoo} for a general review on the topic).

The primary aspect of these developments is that massless gauge fields encompass an infinite set of improper gauge transformations at the boundary of spacetime. These transformations alter the physical state, producing an infinite set of conservation laws expressed as surface integrals. In the quantum theory, these charges are symmetries of the $S-$matrix, which implies the soft theorem as a Ward identity \cite{Strominger:2013jfa}.

This infinite set of conserved quantities has played a key role in identifying soft theorems at both leading \cite{Strominger:2013jfa,He:2014laa,He:2014cra,He:2015zea,Campiglia:2015kxa,Campiglia:2015qka} and subleading \cite{Cachazo:2014fwa,Kapec:2014opa,Lysov:2014csa,Campiglia:2015yka,Campiglia:2016hvg} orders in the energy of the emitted soft particle. More recently, similar symmetry-based arguments have illuminated loop-level corrections as well \cite{Campiglia:2019wxe,Agrawal:2023zea,Donnay:2022hkf,Choi:2024ygx}, suggesting that the infrared perturbative expansion of massless gauge theories can be reorganized through the action of infinite towers of charge operators.

This naturally raises the question of whether the infinite-dimensional symmetry underlying these charges originates from the universal structure of the soft expansion in quantum field theory. In electrodynamics,\footnote{A similar construction of charges for asymptotically flat gravity was first explored in \cite{Conde:2016rom} and later generalized in a broader context in \cite{Freidel:2021dfs,Freidel:2021ytz,Compere:2022zdz}.} these types of charges have also been shown to emerge from the asymptotic expansion of the field equations \cite{Conde:2016csj,Campiglia:2018dyi,Mao:2021eor,Compere:2025tzr}. Their conservation in a scattering process is consistent with diagrammatic results \cite{Hamada:2018vrw,Li:2018gnc}, where the leading and subleading terms correspond to universal soft theorems, while subsubleading terms become non-universal, depending on the specific interaction.

In particular, the results of \cite{Campiglia:2018dyi} show that infinite towers of conserved charges can still emerge from the subleading equations of motion near spatial infinity, independently of the universality of the soft expansion. These include the charges associated with both the leading and subleading soft theorems, along with an infinite extension beyond them that captures subsubleading structures not constrained by universal soft behavior \cite{Bern:2014vva}. Notably, the surface integrals constructed in \cite{Campiglia:2018dyi} offer a notion of conservation that does not necessarily rely on Noether’s theorem, pointing to a broader origin for the underlying symmetry structure.

In this article, we explore the type of conservation law discussed in \cite{Campiglia:2018dyi} in the context of interacting scalar theories. Since these theories lack gauge symmetry, they serve as particularly simple models for studying asymptotic dynamics. Despite their simplicity, they offer valuable insights into how such dynamics constrain the infrared behavior of quantum observables. In particular, it has been shown that the leading soft theorem governing the emission of massless scalars is linked to an infinite-dimensional symmetry arising at the intersection of past and future null infinity \cite{Campiglia:2017xkp,Campiglia:2017dpg}. It is worth emphasizing that the purely massless interacting scalar theory develops a distinct asymptotic structure~\cite{Campiglia:2017xkp}, and to 
our knowledge, an infinite tower of conserved surface charges analogous to the one constructed in this article has not been established in that setting.

We present a systematic procedure to construct an infinite tower of surface integrals that commute with the $S$-matrix of interacting massless scalar fields.  To show this, we notice that the dynamics of the massless scalar obeys the free field equation close to spatial infinity. This simplification permits us to use the results presented in \cite{Henneaux:2018mgn,Fuentealba:2024lll}, where the antipodal conditions relating the dynamics of future and past null infinity have been found for the subleading components of the scalar field \footnote{ The analysis of the asymptotic equations at spatial infinity and its relations with matching conditions at null infinity was first studied in \cite{Troessaert:2017jcm} for gravity. See also  \cite{Capone:2022gme} for an extended treatment that includes non-linearities.}.

Our approach offers a structured way to interpret subleading energy corrections as conservation laws. The surface charges we construct emerge from specific linear combinations of the coefficients in the asymptotic expansion of the massless scalar field. This makes them manifestly finite from the outset, eliminating the need for holographic renormalization.

Crucially, the conservation of this tower of surface integrals arises from a simple but powerful condition: in the absence of News and external sources, the charges remain conserved along retarded (or advanced) time.

To illustrate the implications of these conserved charges, we turn to a scalar version of Yukawa theory, and we analyze the diagrammatic emission of a massless scalar. At leading order in the cubic interaction, we can isolate a contribution to the amplitude that factorizes the hard process from the massless emission. This precise relation can be rederived as a set of subleading soft Ward identities that follow from each conservation law. We focus on the tree-level regime, corresponding to leading order in the coupling. In this approximation, the asymptotic expansion of the scalar field does not develop logarithmic terms, so the tower of charges remains finite and well defined. Beyond tree level, logarithmic corrections are expected to appear and would modify the structure of the charges, a point to which we return in the Discussion \ref{s5}.

The plan of the paper is as follows. In Section \eqref{s2}, we introduce the scalar Yukawa model and derive the subleading soft theorems from a diagrammatic expansion of the scattering amplitude. In Section \eqref{s3}, we analyze the asymptotic dynamics of the massless scalar field and construct an infinite tower of conserved surface charges. We then demonstrate their conservation across spatial infinity and identify their decomposition into soft and hard contributions. We also analyze the matching condition at time-like infinity due to the coupling to massive states. Section \eqref{s4} is devoted to deriving the subleading soft theorems from the conservation of these charges. We provide explicit expressions for the action of the soft and hard charges on the asymptotic states. In Section \eqref{s5}, we discuss the implications of our results, including potential generalizations, and the relevance of logarithmic and loop corrections. Technical details, including properties of the Green’s functions and differential operators on the sphere and hyperboloid, are collected in the Appendix.

\section{Subleading soft expansion for an interacting scalar model}
\label{s2}
We are interested in the amplitude for the emission of a massless scalar $\Phi$. For simplicity, we consider an interaction with a massive field $\Psi$ through a Yukawa-like term,
\begin{align}
\label{L1}
        S[\Phi,\Psi]=\int d^4x\,\left( -\frac{1}{2}( \partial \Phi)^2 -  \left|\partial \Psi \right|^2- M^2  \Psi \Psi^{*}   + g \Psi \Psi^{*}  \Phi \right)\,.
\end{align}
In this section, we derive an infinite set of soft relations using a diagrammatic analysis. In general, a scattering amplitude involving $N$ massive particles (continuous lines) and one massless scalar emission (represented with a dashed line) contains the following Feynman diagrams:
\begin{figure}[h]
\centering
\includegraphics[width=0.5\textwidth]{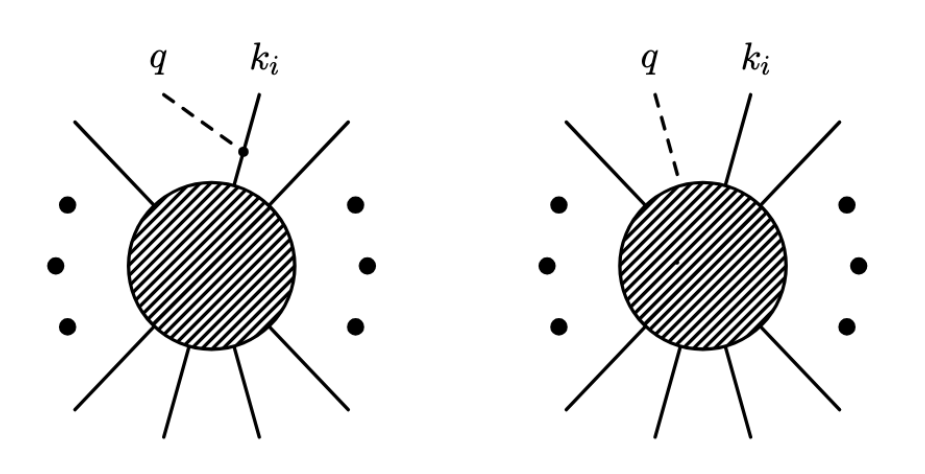}
\end{figure}

Applying the Feynman rules, these diagrams yield 
\begin{align}
    \mathcal{A}_{N+1}(p,k_1,k_2,\dotsc,k_N)= \sum_{j=1}^N \frac{g\, \eta_j}{ 2p \cdot k_j} \mathcal{A}_{N}(k_1,\dotsc,k_j+\eta_j p,...,k_N) + \cN(p;k_1,\dotsc,k_N)\,,
\end{align}
with $\eta_j=1$ for outgoing particles and $\eta_j=-1$ for incoming particles. If the momentum of the massless scalar particle is parametrized as $p=\omega q$, with $q^2=0$, the particle is said to be soft in the limit $\omega \to 0$.  At this point, we assume that the amplitude admits a Laurent expansion in the soft energy 
$\omega$. This is the standard procedure used in the derivation of soft theorems at tree level, where amplitudes are meromorphic functions of external momenta. Beyond tree level, loop corrections are known to generate logarithmic terms in 
$\omega$ (see e.g. \cite{Sahoo:2018lxl}), which would obstruct such an expansion. Since our analysis is restricted to tree level, the Laurent expansion adopted here is consistent and sufficient for deriving the tower of subleading soft relations. Assuming that both functions $\mathcal{A}_N$ and $\cN$ are both analytic in the momentum $p$, we can expand each one of them in a Taylor series around $\omega=0$

\begin{multline}
    \mathcal{A}_{N+1}(\omega\,q,k_1,k_2,\dotsc,k_N)= \sum_{j=1}^N \frac{g\, \eta_j}{ 2 \,\omega\,( q \cdot k_j)} \sum_{l=0}^\infty \eta_j^l\frac{\omega^l}{l!} \left( q\cdot \frac{\partial}{ \partial k_j}\right)^l\mathcal{A}_{N}(k_1,\dotsc,k_N) \\ +\sum_{l=0}^\infty \frac{\omega^l}{l!} \left( q\cdot \frac{\partial}{ \partial p}\right)^l \left.\cN(p;k_1,\dotsc,k_N)\right|_{p=0}\,.\label{diVeccia}
\end{multline}
This construction is similar to that presented in  \cite{Bern:2014vva}, but applied to the scalar case. Notice that the first term is determined to all orders of energy $\omega$ as a differential operator acting on the $N-$point amplitude. The second term contains derivatives of $\cN$ at $q=0$ and represents generic non-factorizable loop order contributions. 

We want to focus on the first term of the expansion \eqref{diVeccia}.
Acting with the operator $ \d^{n}_\omega(\omega \cdot )$ and taking $\omega\to0$ limit, we can single out the $n-th$ term of energy expansion
\begin{align}
    \lim_{\omega \to 0 }\partial^{n}_{\omega}  \left( \omega\mathcal{A}_{N+1}\right) =\sum_{j=1}^N \frac{g \,  {\eta_j^{n+1}}}{ 2 \,( q \cdot k_j)}\left( q\cdot \frac{\partial}{ \partial k_j}\right)^n\mathcal{A}_{N} + n \left( q\cdot \frac{\partial}{ \partial p}\right)^{n-1}\cN(p;k_1,\cdots,k_N) |_{p=0}\,. 
\end{align}
Interestingly, both terms are homogeneous functions of degree $n-1$ in the variable $q$. However, the first term is factorizable because it corresponds to an operator acting on the $N-$point hard amplitude. Furthermore, it contains a single pole at  $q\cdot k_i=0$. 

To project out the $\cN$-term, we will act with a differential operator that removes all the non-factorizable contributions \eqref{diVeccia}. A natural parameterization of the null vector $q^{\mu}$ is 
$q^\mu=(1,\hat{q})$, where $\hat{q}$ is a unit normal vector to the sphere satisfying $\hat{q}^2 =1$. This allows us to express
\be
\label{eqYlm}
\left( q\cdot \frac{\partial}{ \partial p}\right)^l \left.\cN(p;k_1,\cdots,k_N)\right|_{p=0}=\sum^{l}_{k=1} \hat{q}^{i_1}\cdots\hat{q}^{i_k} n_{i_1,\cdots,i_k}\,,
\ee
where $n_{i_1,\cdots,i_k}$ are coefficients depending on the derivatives of $\cN$. But since the product of unitary vectors on the sphere is a (reducible) representation of the spherical harmonics, one can prove that
\be
\dD^{(l+1)}\left(\hat{q}^{i_1}...\hat{q}^{i_l}\right)=0\,, \quad 
\dD^{(n)} \equiv \prod^{n}_{k=1}(D^2 +k(k-1))\,,
\ee
where $D^2$ is the Laplacian over the sphere. A proof of this statement using induction is presented in Appendix \eqref{ops2}.

Applying $\dD^{(n)}$, one finds
\begin{align}
\label{subsfeyn}
  \dD^{(n)} \lim_{\omega \to 0 }\partial^{n}_{\omega}  \left( \omega\mathcal{A}_{N+1}\right) = \dD^{(n)} \left( \sum_{j=1}^N \frac{g \, \eta_j^{n+1}}{ 2 \,( q \cdot k_j)}\left( q\cdot \frac{\partial}{ \partial k_j}\right)^n\mathcal{A}_{N} \right)\,.
\end{align}
It is important to note that due to the simple pole in $q$, this term is not annihilated by the differential operator $\dD^{(n)}$. The previous expression is commonly referred to as the $n$-th subleading soft theorem. In this work, we will show that it follows from the existence of an infinite tower of surface integrals that commute with the $S$-matrix.

To clarify potential ambiguities in \eqref{subsfeyn}, observe that the operator $q\cdot \frac{\partial}{\partial k_j}$ does not preserve the mass-shell condition of the hard external particles. However, the subsequent action of $\dD^{(n)}$ eliminates any discrepancy arising from this fact, ensuring that equation \eqref{subsfeyn} holds on-shell. Moreover, we follow the same prescription as in \cite{Cheung:2021yog}, where the $N$-point amplitude $\mathcal{A}_N$ is taken to be on-shell. Accordingly, the subleading soft relations are valid when one first uses momentum conservation to eliminate one of the hard momenta and then takes the soft limit $\omega \to 0$.

\section{Charge construction}
\label{s3}

In this section, we demonstrate that an infinite tower of charges,
describing the subleading soft scalar theorems at any order in frequency,
can be derived by solving the wave equation in the presence of a source. This derivation relies on a series expansion in inverse powers of the radial distance
of the scalar field $\phi$. To construct the subleading charges,
we will impose the following condition: we consider a linear combination
of the various orders of the scalar field and require that, in the absence of scalar radiation
and hard fluxes across the boundary, the resulting
expressions do not change under evolution in retarded time. This condition uniquely determines
both the soft and hard charges.

 Before proceeding with the explicit construction, let us clarify the motivation for the radial expansion in relation to the conservation due to the matching conditions across $i^0$. 
The scalar theory under consideration does not possess gauge symmetry, and therefore the origin of the conserved quantities is not related to residual gauge transformations. Instead, the key ingredient is the asymptotic free-field behavior of the scalar near spatial infinity $i^0$. This simplification arises due to the massive origin of the sources considered in this work. 

From the Hamiltonian point of view, the profile of the field near $i^0$ provides initial data organized in a $1/r$ expansion \cite{Henneaux:2018mgn, Fuentealba:2024lll}. These data evolve under the free wave equation in the asymptotic region and their structure is preserved under time evolution. This allows one to establish matching conditions between $\mathscr I^-_+$ and $\mathscr I^+_-$, which in turn guarantee the conservation of the associated surface integrals throughout the scattering process. We will make an explicit connection to this point in Section \ref{HamiltonianMO}.

\subsection{Radial expansion of the massless scalar field in retarded null coordinates}
Let us express the equations of motion for the massless scalar field
$\phi\left(x\right)$ in the presence of an external source $J$. By using retarded null coordinates $\left(u,r,x^{A}\right)$,
which are related to Minkowski coordinates $x^{\mu}$ by
\[
x^{0}=u+r\,,\qquad\qquad x^{i}=r\hat{n}^{i}.
\]
Here, $x^{A}$ with $A=1,2$  represents coordinates on the two-sphere,
and $\hat{n}$ is the unit vector normal to the two-sphere. In
spherical coordinates, this vector is given by $\hat{n}=\left(\sin\theta\cos\varphi,\sin\theta\sin\varphi,\cos\theta\right)$. 

In retarded null coordinates, the wave equation 
\be 
\label{eq1.1}
\square\phi=-J,
\ee
takes the form,
\begin{equation}
-2\partial_{u}\partial_{r}\phi+\partial_{r}^{2}\phi+\frac{2\partial_{r}\phi}{r}-\frac{2\partial_{u}\phi}{r}+\frac{1}{r^{2}}D^{2}\phi=-J.\label{eq:field_equation_massless}
\end{equation}
We model the effect of the interaction by introducing a generic external source 
$J$. This simplification is sufficient to capture the tree-level dynamics relevant for the derivation of the subleading soft charges. Indeed, in the Yukawa theory considered here, the current $J$ is generated by the massive field, but to leading order in the coupling, its backreaction does not alter the free propagation of the massless scalar near null infinity. In this approximation, the field equations reduce to those of a free scalar sourced linearly, which is precisely the regime where our construction of conserved charges applies. Nonlinear effects, such as those responsible for logarithmic terms in the asymptotic expansion, only appear beyond tree level and therefore lie outside the scope of the present work.

Let us consider the expansion of the scalar field and the current in a series in inverse
powers of the radial coordinate:
\be
\label{eq1.2}
\phi\left(x\right)=\sum_{k=0}\frac{\phi^{\left(k\right)}\left(u,x^{A}\right)}{r^{k+1}}\,,\qquad\qquad \left.J\right|_{\dI^+}=\sum_{k=0}\frac{J^{\left(k\right)}\left(u,x^{A}\right)}{r^{k+3}}.
\ee
Substituing this expansion into the field equation (\ref{eq:field_equation_massless}),
one obtains the following recursion relation \cite{ Satishchandran:2019pyc,Bekaert:2022ipg}
\begin{equation}
\partial_{u}\phi^{\left(k+1\right)}= -\frac{1}{2\left(k+1\right)}\left[D^{2}+k\left(k+1\right)\right]\phi^{\left(k\right)}+\frac{1}{2\left(k+1\right)}J^{\left(k\right)},\label{eq:rec_rel}
\end{equation}
for $k=0,1,2,\dots$.  It is important to note that $\phi^{\left(0\right)}\left(u,x^{A}\right)$ is not determined by the field equations; rather, it constitutes part of the free data. Consequently, the radiation field at infinity is encoded in its time dependence. In other words, the scalar Bondi News are given by $\partial_{u}\phi^{\left(0\right)}$. This can also be seen from the expression for the energy flux across null infinity \cite{Frolov:1977bp}
\[
\Delta E=\lim_{r\to \infty}\int_{\dI^+}d^3\Sigma^{\mu} \,T^{\phi}_{\mu u}=\int_{\dI^{+}}dud^{2}x\sqrt{\gamma}\left(\partial_{u}\phi^{\left(0\right)}\right)^{2}\,,
\]
that has been derived from the projection of the scalar stress-energy tensor on the $u={\rm const}$ surface. Here $\gamma$ denotes the determinant of the metric of the round 2-sphere $\gamma_{AB}$. This expression is the scalar field analog of Bondi's energy loss formula, indicating that the energy flux vanishes if and only if  $\partial_{u}\phi^{\left(0\right)}=0$.

\subsection{Infinite tower of charges}
We turn to the construction of the infinite tower of conserved charges at future null infinity $\dI^+$ that will be associated with the emission of a massless scalar. At the end, we comment on the suitable modifications of this construction for the derivation of the same charge at past null infinity $\dI^-$.

As an ansatz for the 
$m$-th charge, we propose the following linear combination of the various orders of the scalar field
\begin{equation}
Q_{\left(m\right)}:=\sum_{k=0}^{m}\oint_{S^{2}} d^{2}x\sqrt{\gamma}\,\mu^{+}_{\left(k\right)}\phi^{\left(k\right)},\label{eq:Q_ansatz}
\end{equation}
where the integral is taken over the celestial sphere. The case $m=0$ represents the charge used in  \cite{Campiglia:2017xkp} to obtain the leading scalar soft theorem.

The coefficients $\mu^+_{(k)}$ are determined by imposing the following condition: the charge $Q_{(m)}$ must be conserved under the continuous evolution in retarded time, i.e., $dQ_{(m)}/du=0$ when there are no News ($\dot{\phi}^{(0)} = 0$) and external currents ($J = 0$) are absent. 

By applying the recursion relation \eqref{eq:rec_rel}, it is straightforward to demonstrate that in the absence of News and currents, the $u$-derivative of $Q_{(m)}$ is given by:
\[
\partial_{u}Q_{\left(m\right)}=\oint_{S^{2}}d^{2}x\sqrt{\gamma}\left[\left(\partial_{u}\mu^+_{\left(m\right)}\right)\phi^{\left(m\right)}+\sum_{k=0}^{m-1}\left(\partial_{u}\mu^+_{\left(k\right)}-\frac{\mu^+_{\left(k+1\right)}}{2\left(k+1\right)}\left[D^{2}+k\left(k+1\right)\right]\right)\phi^{\left(k\right)}\right]=0.
\]
Consequently, the above condition uniquely determines $\mu^+_{(k)}$ for any given values of $\phi^{(k)}$ through the following recursion relation:
\begin{equation}
\partial_{u}\mu^+_{\left(k\right)}=\frac{1}{2\left(k+1\right)}\left[D^{2}+k\left(k+1\right)\right]\mu^+_{\left(k+1\right)},\label{eq:rec_mu}
\end{equation}
for $k=0,1,\dotsc,m-1$, and where $\partial_{u}\mu_{\left(m\right)}=0$.
The conserved charges can be explicitly constructed by integrating the preceding equations and substituting the results into Eq. \eqref{eq:Q_ansatz}. The general solution is a linear combination of $m$ functions on the sphere. For our purposes, it will only be necessary to work with the particular solution for $\mu^+_{(k)}$, which can be expressed as
\be
\label{par_f} 
\mu^+_{(k)}(u,\hat{n})= \left(\frac{u}{2}\right)^{m-k} \frac{k!}{(m-k)! \;m!} \left(\prod_{i=1}^{m-k}\left(D^{2}+\left(m-i\right)\left(m-i+1\right)\right)\right) f_+(\hat{n})\,,
\ee
that contains a power dependence on the retarded time and is also parametrized in terms of a single arbitrary parameter on the sphere, denoted by $f_+$. This relation works for $k\in \{ 1,...,m\}$, and $\mu_m=f_+(\hat{n})$.
Plugging back this parameter in the charges on finds the following general expression for the $m$-th charge:
\be
\label{Qms2}
Q_{(m)}\left[f_+\right]=\oint_{S^{2}} d^{2}x\sqrt{\gamma}f_+\left(\hat{n}\right) \left( \phi^{(m)}+\sum_{k=1}^{m}\frac{\left(m-k\right)!}{2^{k}m!k!}u^{k}\left[\prod_{n=1}^{k}\left(D^{2}+\left(m-n\right)\left(m-n+1\right)\right)\right]\phi^{\left(m-k\right)}\right).
\ee
In order to visualize the dependence on differential operators, we explicitly write down the first four charges 
\begin{align*}
Q_{(0)}\left[f_+\right] & =\oint d^{2}x\sqrt{\gamma}\,f_+\left(\hat{n}\right)\phi^{\left(0\right)}\\
Q_{(1)}\left[f_+\right] & =\oint d^{2}x\sqrt{\gamma}\,f_+\left(\hat{n}\right)\left(\phi^{\left(1\right)}+\frac{u}{2}D^{2}\phi^{\left(0\right)}\right),\\
Q_{(2)}\left[f_+\right] & =\oint d^{2}x\sqrt{\gamma}\,f_+\left(\hat{n}\right)\left[\phi^{\left(2\right)}+\frac{u}{4}\left(D^{2}+2\right)\phi^{\left(1\right)}+\frac{u^{2}}{16}D^{2}\left(D^{2}+2\right)\phi^{\left(0\right)}\right],\\
Q_{(3)}\left[f_+\right] & =\oint d^{2}x\sqrt{\gamma}\,f\left(\hat{n}\right)\left[\phi^{\left(3\right)}+\frac{u}{6}\left(D^{2}+6\right)\phi^{\left(2\right)}+\frac{u^{2}}{48}\left(D^{2}+2\right)\left(D^{2}+6\right)\phi^{\left(1\right)}\right.\\
&\qquad\qquad\qquad\qquad\qquad\qquad\qquad\qquad\left.+\frac{u^{3}}{288}D^{2}\left(D^{2}+2\right)\left(D^{2}+6\right)\phi^{\left(0\right)}\right],\\
 &\qquad\qquad\qquad\qquad\qquad\qquad\qquad \vdots
\end{align*}
where we can see the emergence of differential operators that annihilate higher-order spherical harmonics. It can also be noticed that the first elements of this tower of charges resemble the results found in electromagnetism for the subleading soft theorems \cite{Campiglia:2016hvg, Campiglia:2018dyi}, where the role of $\phi^{(k)}$ is played by the subleading orders of the radial electric field $F_{ur}^{(k)}$.

So far we have defined the surface integral \eqref{Qms2} on any sphere at null infinity. Considering this expression, it is natural to wonder about the finiteness of $Q_{(m)}$ in the limit $u\to \pm\infty$. To demonstrate that, it is essential to show that our expansion contains no $\log(u)$ terms in it.  
Indeed, let us assume that the field $\phi^{(k)}$ can be expanded as
\begin{align}
    \phi^{(k)}(u,x^A)=\sum_{m\in \mathbb{Z}} \frac{1}{u^m}\phi^{(k,m)}(x^{A})\,.
\end{align}
In this notation, $\phi^{(k,m)}$ is the term that goes as $u^{-m}$ and $r^{-k-1}$. Replacing this expansion in \eqref{eq:rec_rel}, we find a particular subsidiary condition from the order $O\left( 1/u\right)$,
\be
    (D^2 + k(k+1))\phi^{(k,1)}(x^{A})=0\,.
\ee
Since the massless field interacts with massive particles, one can assume that the external current $J$ does not have support on $\dI^{\pm}$, so the additional terms proportional to the current can be neglected.
Applying this equation to \eqref{eq:rec_rel} ensures that, after integrating in $u$, the term proportional to $\log(u)$ will vanish. Furthermore,  integrating  order by order the sequence given in \eqref{eq:rec_rel},  we obtain a general expansion   for $\phi^{(k)}$
\begin{align}
\label{eqphim}
    \phi^{(k)}(u,x^{A})= \phi^{(k,0)}(x^{A}) +\sum_{p=1}^k (-1)^{p} \frac{u^{p}(k-p)!}{2^{p}\,p! \,k! } \prod_{l=k-p}^{k-1}(D^2 + l(l+1)) \phi^{(k-p,0)}(x^{A}) + O\left( \frac{1}{u}\right).
\end{align}
Combining this formula with \eqref{Qms2}, we obtain the simplified version of the charge evaluated at $u \to - \infty $
\begin{align}
    Q_{(m)}[f]= \oint_{\dI^+_-} d^2x \sqrt{\gamma} \, f_+(x^{A}) \phi^{(m,0)}(x^{A})\,,
\end{align}
demonstrating that it is finite as $u$ approaches $-\infty$. It is important to emphasize that the same  $u$-independent term will be obtained by performing an expansion as $u$ approaches $+\infty$.

\paragraph{Past null infinity analysis:} The analysis presented before can be replicated at $\dI^{-}$, by noticing that the recursion relation \eqref{eq:rec_rel} changes to
\be
\partial_{v}\phi^{\left(k+1\right)}=+\frac{1}{2\left(k+1\right)}\left[D^{2}+k\left(k+1\right)\right]\phi^{\left(k\right)}+\frac{1}{2\left(k+1\right)}J^{\left(k\right)},\label{eq:rec_rel2}
\ee
where $v=t+r$ is the advanced time. This implies that the condition on the parameter $\mu^-_{(k)}$ will differ by a sign from the one obtained in  \eqref{eq:rec_mu}. Therefore, the corresponding particular solution for this parameter will be given by 
\be
\label{par_fv} 
\mu^-_{(k)}(v,\hat{n})= \left(-\frac{v}{2}\right)^{m-k} \frac{k!}{(m-k)! \;m!} \left(\prod_{i=1}^{m-k}\left(D^{2}+\left(m-i\right)\left(m-i+1\right)\right)\right) f_-(\hat{n})\,.
\ee
In what follows we will determine the matching condition between the parameters $f_+$ and $f_-$ so that the charge constructed here is conserved in scattering processes.

\subsection{Infinite tower of conservation laws}
\label{HamiltonianMO}
The charges discussed in the previous section suggest the existence of an infinite set of conservation laws. This can be shown by carefully examining the field equation for $\phi$ in the vicinity of spatial infinity $i^0$. Our approach closely follows the analysis presented in \cite{Henneaux:2018mgn, Fuentealba:2024lll}, where the general solution has been studied and constructed in detail.

The wave equation in the neighborhood of $i^0$ provides a smooth limit that connects quantities defined on the past boundary of future null infinity, $\dI^{+}_{-}$ (as $u \to -\infty$), with those on the future boundary of past null infinity, $\dI^{-}_{+}$ (as $v \to +\infty$). Using this correspondence, we will establish the following conservation law:
\be
\label{conservation}
\left. Q^{+}_{(m)} [f_+] \right|_{\dI^{+}_{-}}=\left. Q^{-}_{(m)}[f_-] \right|_{\dI^{-}_{+}}\,.
\ee
For this equality to hold, we must assume that the parameter $f$ is antipodally related on the sphere, following the condition
\be
\label{anti}
 f_+(\hat{x})=(-1)^{m} f_-(-\hat{x})\,,
\ee
where $\hat{x}=(\theta,\phi)$, and the notation $-\hat{x}$ indicates the antipodal transformation $-\hat{x}=(\pi-\theta, \phi+\pi)$. It is important to clarify the role of the function 
$f_{\pm}$ in this construction. The freedom to choose 
$f_{\pm}$ does not introduce an additional independent assumption, nor a new dynamical input beyond the matching condition itself. Rather, $f_{\pm}$ plays the role of a smearing function for the underlying angle-dependent charge density defined at the boundary. Equivalently, one may expand this parameter in spherical harmonics and regard the conserved quantities $Q_{(m)}[Y_{\ell m}]$ as a basis of charges labelled by angular modes on the sphere. In this sense, the family of charges is the harmonic decomposition of the same angle-dependent conservation law implied by the matching conditions.

We recall that $m$-th component of the field evaluated on $u\to-\infty$ reduces to 
\be
Q^{+}_{(m)}[f_{+}]= \oint_{\dI^+_-} d^2x \, \sqrt{\gamma} f_+(\hat{x})\phi^{(m,0)}(\hat{x})\,.
\ee
A similar evaluation close to $\dI^-_+$, yields the expansion in the advanced time for $\phi^{(m)}(v,x)$, giving
\be
Q^{-}_{(m)}[f_{-}]= \oint_{\dI^-_+} d^2x \, \sqrt{\gamma} f_-(\hat{x})\phi^{(m,0)}(\hat{x})\,.
\ee
We shall examine the wave equation near spatial infinity to demonstrate that the two surface integrals discussed above are identical, provided \eqref{anti}. We start by reviewing some of the results presented in \cite{Henneaux:2018mgn,Fuentealba:2024lll}. Hyperbolic coordinates are used to foliate Minkowski spacetime close to space-like infinity:
\be
\eta=\sqrt{r^2-t^2}\,, \quad s=\frac{t}{r}\,.
\ee
In these coordinates, the scalar field admits the decomposition
\be
\label{scai0}
\phi(\eta,s,\hat{x})=\sum^{\infty}_{k=0}\sum_{l,m}(1-s^2)^{\tfrac{1-k}{2}}\eta^{-(k+1)} \Theta^{(k)}_{l m}(s) Y_{lm}(\hat{x})\,,
\ee
which reduces $\Box \phi=0$ to a linear second order differential equation for the functions $\Theta^{(k)}_{lm}$,
\be
\label{eqMCO}
(1-s^2)\d^2_s \Theta^{(k)}_{lm}+2(k-1) s \d_s \Theta^{(k)}_{lm} +\left[l(l+1)-k(k-1)\right] \Theta^{(k)}_{lm} =0 \,,
\ee
that for $k=0$ corresponds to Legendre’s equation. The generalization for $0<k \leq l$ can be expressed in terms of the Gegenbauer's polynomials $P^{\lambda}_{n}(s)$ \cite{Szego1975}. In this case, the solution reads
\be
\Theta^{(k)}_{lm}(s)=(1-s^2)^k P^{\left(k+1/2\right)}_{l-k}(s)\,.
\ee
There is also a branch of solutions for $k > l$, given by polynomial solutions. Details on this specific form of this family of solutions can be found in Appendix A of 
\cite{Henneaux:2018mgn}. Following their conventions, we will denote these solutions as 
$\Theta^{(k)}_{lm}(s)\equiv\tilde{P}^{\left(k+1/2\right)}_{l-k}(s).$

It is important to note that a separate class of solutions to \eqref{eqMCO} contains logarithmic terms. However, such solutions are excluded from our discussion under the boundary conditions assumed here, specifically \eqref{eq1.2} at $\mathscr{I}^\pm$.

We would like to read off the component $\phi^{(k,0)}$ from the solution \eqref{scai0}. Expressing hyperbolic coordinates in terms of retarded time $u=t-r$, we find
\begin{multline}
\label{expansionphi}
\phi(u,r,\hat{x})=\sum^{\infty}_{k=0}\left(\sum_{l \geq k, m} r^{-k-1} \Phi^{k}_{lm} P^{\left(k+1/2\right)}_{l-k}\left(1+u/r\right) Y_{lm}(\hat{x}) \right. \\ \left.+\sum_{l<k, m} r^{-1}(-2u)^{-k} \left(1+\frac{u}{2r}\right)^{-k}  \Phi^{k}_{lm} \tilde{P}^{\left(k+1/2\right)}_{l-k}\left(1+u/r\right) Y_{lm}(\hat{x})\right)\,,
\end{multline}
where $\Phi^{k}_{lm}$ are integration constants. Let us note that terms in the sum with $l<k$ have an explicit dependence on $u$. However, upon expanding the full expression in inverse powers of $r$, these terms can also contribute to the $u$-independent coefficient at order $r^{-(k+1)}$. Comparing the above expansion with \eqref{eqphim} and \eqref{eq1.2}, one finds
\be
\label{changed}
\phi^{(k,0)}_{\dI^+_-}(x)= \sum_{l \geq k, m} \Phi^{k}_{lm} P^{\left(k+1/2\right)}_{l-k}(1) Y_{lm}(\hat{x})
+\sum_{l<k, m}  \Phi^{k}_{lm}\, c^{(k)}_{l}\, Y_{lm}(\hat{x})\,,
\ee
where $c_l^{(k)}$ is the coefficient of the $u$-independent term at order $r^{-(k+1)}$ arising from the large-$r$ expansion of the $\tilde{P}$-branch. Analogously, using the advanced time $v$, the $v$-independent component of the scalar field reads
\be
\phi^{(k,0)}_{\dI^-_+}(x)= \sum_{l \geq k, m} \Phi^{k}_{lm} P^{\left(k+1/2\right)}_{l-k}(-1) Y_{lm}(\hat{x})
+\sum_{l<k, m}  \Phi^{k}_{lm}\, d^{(k)}_{l}\,Y_{lm}(\hat{x})\,.
\ee
As shown in Appendix~\ref{app:polybranch_parity}, the coefficients defined above satisfy
\be
\label{nueva}
d_l^{(k)} = (-1)^{\,l-k} c_l^{(k)}\,.
\ee
Furthermore, combining the parity relation for the Gegenbauer polynomials \cite{Fuentealba:2024lll},
\be
P_{m}^{(k)}(-1)=(-1)^{m} P_{m}^{(k)}(1)\,,
\ee
with the spherical-harmonic identity $Y_{lm}(-\hat x)=(-1)^l Y_{lm}(\hat x)$
and Eq.~\eqref{nueva}, one finds
\be
\phi^{(k,0)}_{\dI^+_-}(-\hat{x})= (-1)^{k} \phi^{(k,0)}_{\dI^-_+}(\hat{x})\,,
\ee
which shows that the charges $Q_{(m)}[f]$ are indeed conserved \eqref{conservation}.

\paragraph{Transformation properties under Poincaré: }
Since the asymptotic conditions for the scalar field \eqref{eq1.2} are invariant under Poincaré symmetry, we analyze the transformation properties of the field $\phi^{(n,0)}(x)$ that is central in the charges constructed before. Infinitesimally, these transformations are generated by vector fields $\xi$ whose components in retarded coordinates are given by\footnote{A similar expression is obtained in advanced coordinates using the matching condition.}
\be  \xi^{u}=\alpha(\hat{x}) + \frac{u}{2} D_A Y^A(\hat{x})\,, \quad \xi^{r}=-\frac{1}{2}(u+r)D_A Y^A(\hat{x})+\frac{1}{2}D^2\alpha\,, \quad \xi^A=Y^A-\frac{1}{r}\gamma^{AB} D_B \xi^{u}\,, \ee
where translations are represented by the function $\alpha$, and Lorentz transformations are encoded in the vector fields $Y^A$ on the sphere. A useful decomposition is given by
\be \label{Hem} \alpha(\hat{n})=-\alpha_0+\vec{\alpha}\cdot \hat{n}\,, \quad Y^{A}=D^A (\vec{\beta}\cdot\hat{n}) +\frac{1}{\sqrt{\gamma}}\epsilon^{AB}D_B (\vec{\Omega}\cdot\hat{n})\,. \ee
Here, $\alpha_0$ and $\vec{\alpha}$ correspond to global time and spatial translations, respectively. The vector $Y^A$ is expressed into boosts, represented by gradients of the unit normal vector $\hat{n}$, and rotations, characterized by the constant vector $\vec{\Omega}$ acting through the curl of $\hat{n}$.

Computing the Lie derivative of the scalar field gives the transformation laws under translations and Lorentz transformations. One finds that the $u$-independent part of the field transforms as
\be 
\begin{split} 
\delta_\alpha \phi^{(n,0)}&= \alpha \partial_u \phi^{(n,0)}+ (n+1)(\vec{\alpha}\cdot \hat{n}) \phi^{(n-1,0)}+ \partial^A(\vec{\alpha}\cdot \hat{n})\partial_A \phi^{(n-1,0)}\,,\\ 
\delta_Y \phi^{(n,0)}&=Y^A \partial_A \phi^{(n,0)}+\frac{(n+1)}{2} D_A Y^A \phi^{(n,0)}\,. \end{split} 
\ee
Under global translations, the field mixes with the preceding order in the radial expansion of $\phi$. Under Lorentz transformations, the decomposition \eqref{Hem} shows that $\phi^{(m,0)}$ transforms as a scalar under rotations, while under boosts it behaves as a density of weight $(n+1)/2$.

From the perspective of the Celestial holography program (see e.g. \cite{Raclariu:2021zjz,Pasterski:2021raf} for a review on the subject), one can characterize the field $\phi^{(n,0)}$ using coordinates such that $\gamma_{AB}dx^A dx^B=2 dz d\bar{z}$ is locally flat. The action under boosts then becomes
\be 
\label{deltaY}
\delta_Y \phi^{(n,0)}= Y^z \partial_z \phi^{(n,0)} +Y^{\bar{z}} \partial_{\bar{z}} \phi^{(n,0)}+\frac{(n+1)}{2}\left( \partial_z Y^z+ \partial_{\bar{z}} Y^{\bar{z}} \right) \phi^{(n,0)}\,, \ee
demonstrating that the field $\phi^{(n,0)}$ is a $(\tfrac{n+1}{2},\tfrac{n+1}{2})$ primary operator under the global ${\rm SL}(2,\mathbb{C})$ conformal group in two dimensions.

\subsection{Soft and hard charges}

Note that the tower of charges in Eq. \eqref{Qms2} is defined at any retarded time $u$. In particular, they are conserved under the evolution in $u$ in the absence of News and external currents. However, to make contact with the soft theorems, one must account for non-trivial fluxes across the boundary. 

Following Strominger \cite {Strominger:2017zoo}, it is convenient to define the charge at $\dI^{+}_{-}$ (the limit as $u \to -\infty$ in $\dI^+$) and allow for non-vanishing radiation across the boundary. Thus, the relevant charge is given by:
\be
\label{eq:Tower_charges}
Q_{(m)}\left[f_+\right]=\lim_{u\to-\infty }\oint_{S^2} d^{2}x\sqrt{\gamma}f_+\left(\hat{x}\right) \left( \phi^{(m)}+\sum_{k=1}^{m}\frac{\left(m-k\right)!}{2^{k}m!k!}u^{k}\left[\prod_{n=1}^{k}\left(D^{2}+\left(m-n\right)\left(m-n+1\right)\right)\right]\phi^{\left(m-k\right)}\right).
\ee
A similar construction can be straightforwardly implemented at past null infinity $\dI^{-}$ to define the charges $Q^{-}_{(m)}\left[f\right]$. Let us focus on \eqref{eq:Tower_charges} for concreteness.

By applying Stoke's theorem,  the charge \eqref{eq:Tower_charges} can be decomposed into two contributions: one evaluated at $\dI^{+}$ and the other at $\dI^{+}_{+}$ (the limit $u \to +\infty$ of $\dI^{+}$). Furthermore, we apply the recursion relation \eqref{eq:rec_rel}, under the assumption that the currents $J^{(k)}$ vanish fast enough close to $\dI^{+}$. This assumption holds because we are considering massive hard particles that never reach $\dI^{+}$. Thus, one obtains
\[
Q_{(m)}\left[f_+\right]=Q_{m}^{\text{soft}}\left[f_+\right]+Q_{m}^{\text{hard}}\left[f_+\right],
\]
with\footnote{We define the operator  $\dD^{(0)}$ as the identity}
\begin{align}
\label{qS}
Q_{m}^{\text{soft}}\left[f_+\right]&=-\frac{1}{2^{m}\left(m!\right)^{2}}\int_{\dI^+}d^{2}xdu\sqrt{\gamma}\dD^{(m)} f_+ \,\, u^{m}\partial_{u}\phi^{\left(0\right)}\,,\\ \label{QH}
Q_{m}^{\text{hard}}\left[f_+\right]&=\oint_{\dI_{+}^{+}}d^{2}x\sqrt{\gamma}\,f_+ \phi^{\left(m\right)}_{\dI^+_+} \,.
\end{align}
Here, we use the term ``soft charge'' to refer to the piece of the charge that depends solely on $\partial_{u}\phi^{(0)}$ and is integrated over $\dI^{+}$. However, this terminology should not cause confusion, as these charges are not necessarily associated with massless scalar particles of very low energy. As we will demonstrate in the following sections, the charges for $m \geq 1$ are associated with subleading soft theorems, which generally involve higher values of the frequency.

To derive the $m$-th subleading soft theorem from the conservation of the $m$-th charge \eqref{eq:Tower_charges}, it is necessary to simplify the expression for  $Q_{m}^{\text{hard}}$, and then express it terms of the creation and annihilation operators of massive particles. This analysis will be addressed in the next section, by studying the behavior of a massless scalar field sourced by an external massive current.

\subsection{Asymptotic dynamics at $i^+$}
The purpose of this section is to provide an expression for the field $\phi$ close to time-like infinity compatible with the asymptotic behavior displayed at the future of $\dI^+$, represented by \eqref{eq1.2}. Since this region of the conformal diagram is only reached by massive fields, we  will analyze the dynamics of a massless scalar field sourced by $J$ through the equation \eqref{eq1.1}, where the external interaction admits an expansion around time-like infinity $i^+$  given by
\be
J=\sum^{\infty}_{n=0}\frac{J_{n}(Y)}{\tau^{n+3}}\,,
\ee
where $\tau$ is a time-like coordinate that parameterizes the hyperboloids $x^\mu x_\mu=-\tau^2$. This relation means that we can choose embedding coordinates such that 
\be
\label{Ydef}
x^{\mu}(\tau,\rho,\hat{x})\equiv\tau Y^{\mu}=\tau(\sqrt{\rho^2+1},\rho\,\hat{n})\,,
\ee
where $\tau,\rho>0$, with $Y$ a time-like vector normalized by $Y \cdot Y =-1$. The Minkowski line element then reads
\be
ds^2=-d\tau^2+\tau^2\left( \frac{d\rho^2}{1+\rho^2} +\rho^2 \gamma_{AB}dx^A dx^B\right)\,.
\ee
We can now solve the equation \eqref{eq1.1} by assuming the following expansion of the field
\be
\phi(x)=\sum^{\infty}_{n=0}\frac{\phi_n(Y)}{\tau^{n+1}}\,,
\ee
obtaining
\be
\left(\Delta -(n^2-1)\right)\phi_n(Y)=-J_{n}(Y)\,.
\ee
We can express the solution in terms of the Green function on the hyperboloid, finding
\be
\label{GH}
\phi_n(Y)=\int d^3V' \, \cG_n(Y,Y') J_n(Y')\,, \quad {\rm with} \quad \left(\Delta -(n^2-1)\right)\cG_n(Y,Y')=-\frac{1}{\sqrt{h}}\delta^{(3)}(y-y')\,,
\ee
we note that $y^a=(\rho, x^A)$ are coordinates embedded in the hyperbolic plane and $\Delta$ denotes the corresponding Laplacian operator.

Since this manifold is a homogeneous space, the solution $\cG_n$ must be a function of $|Y-Y'|^2$, or equivalently, of $\sigma\equiv-Y\cdot Y'$. Solving the homogeneous equation in terms of $\sigma$ and imposing the presence of the source condition at $Y=Y'$, we find
\be
\cG_n(Y,Y'):=\frac{1}{4\pi } \frac{1}{\sqrt{\sigma^2-1} (\sigma + \sqrt{\sigma^2-1})^n}\,.
\ee
In finding the above solution, we have chosen the branch that vanishes when $\sigma\to \infty$. 

We can now  map the information at $i^+$, provided by the sources $J_n(Y)$, with the coefficients $\phi^{(n)}$  at $\dI^+_+$. To do so, we use the relation between the retarded Bondi and the hyperbolic coordinates
\be
u=\tau(\sqrt{1+\rho^2}-\rho)\,, \quad r=\tau\rho\,.
\ee
In particular, one can see that $\tau \sigma = - u (Y'^0) + r (q\cdot Y')$ where $Y'^0=\sqrt{1+\rho'^2}$ and $q=(1,\hat{q})$ a null vector. Furthermore, it is possible to show that for large $r$, we find 
\be
\frac{\cG_n(Y,Y')}{\tau^{n+1}}= \frac{1}{2\pi \, r^{n+1}} \frac{1}{(-2Y\cdot q)^{n+1}} +\sum^{\infty}_{k=1} \frac{u^k}{r^{n+k+1}} \cR_{(k)}(q, Y')\,, 
\ee
with $\cR_{(k)}$ terms that can be found explicitly, although their exact forms are not crucial to the discussion here. The main observation is that the matching condition
$$\left. \phi\right|_{\dI^+_+}=\left. \phi \right|_{i^+}\,,$$
yields the coefficient of the asymptotic expansion \eqref{eq1.2} 
\be
\left. \phi^{(n,0)}(x)\right|_{i^+}=\int d^3 V\left[\frac{J_{n}(Y)}{2\pi (-2 Y\cdot q)^{n+1}}\right ]\,.
\ee
This value is necessary to evaluate the hard charge found in the previous section \eqref{QH}. The expression for $Q^{(\rm hard)}_m$ renders the finite result
\begin{align}
\label{hardip}
Q_{m}^{\text{hard}}\left[f_{m}\right]&=\int_{i^+} d^3V \,  f_{m} (Y;q) \, J_{m}(Y)\,,\\
\label{hardip2}
f_{m} (Y;q)&=\frac{1}{2\pi} \oint d^{2}x \sqrt{\gamma} \frac{f(\hat{x})}{(-2 q(\hat{x})\cdot Y)^{m+1}}\,.
\end{align}
For $m=0$ this expression agrees with the leading hard charge found 
in \cite{Campiglia:2016hvg}. For $m>0$, the above provides a natural generalization consisting of sub-leading components of the current smeared with the extension on $i^+$ of the parameters $f$ defined at null infinity. In fact, the expression
\be
G_{m}(Y,q):= \frac{\sqrt{\gamma}}{(-2 Y\cdot q)^{m+1}}\,,
\ee
corresponds to the null to time-like infinity scalar propagator defined in \cite{Campiglia:2015lxa}. Furthermore, it satisfies
\be
\label{a5}
\Delta G_{m}= (m^2-1) G_m\,,
\ee
that can be checked by direct differentiation and using the properties outlined in the appendix \eqref{aaYY}. It follows that $f_m(Y)$ also satisfies the same differential equation.
\
\subsection{Subleading hard current from the interacting model}
\label{sub-mass}
To study the consequence of the hard charge at the quantum level, it is necessary to construct a basis of massive oscillators producing the subleading currents $J_m(Y)$. Here, we will find a representation of those currents by solving the leading equations iteratively inside the bulk.

We are interested in the fully interacting Lagrangian coupling a real massless field through a Yukawa type of interaction \eqref{L1}. The corresponding field equations read
\be
\label{eqM}
\Box \Phi=-g \Psi \Psi^{*}\,, \quad
(\Box -M^2)\Psi=-g \Phi \Psi\,,
\ee
together with the equation for the complex conjugate $\Psi^*$ that is readily obtained from conjugating the last equation. Expanding the equations around the free solutions, 
\be
\Phi=\phi+O(g),\quad \Psi=\psi+O(g)\,.
\ee
It is worth emphasizing that this approximation is enough for studying the subleading soft theorems analyzed in this article. In this regime, we identify that $\psi$ obeys the massive free equation and we obtain the value of $J$ in \eqref{eq1.1}
\be
(\Box-M^2)\psi=0\,, \quad J=g\psi \psi^{*}\,.
\ee

The equation for $\psi$ can be solved via a series expansion around future time-like infinity as follows:
\begin{equation}
\psi(x) = e^{\alpha \tau} \sum_{n=0}^{\infty} \frac{\psi_{(n)}(Y)}{\tau^{n+3/2}}.
\end{equation}
The coefficients $\psi_{(n)}$ satisfy the recurrence relation:
\begin{equation}
\psi_{n+1} = \frac{1}{2\alpha(n+1)} \left[-\Delta + (n+3/2)(n-1/2)\right] \psi_{n},
\end{equation}
where $\alpha^2 = -M^2$.

The above recurrence relation allows for the determination of all $\psi_{n}$ with $n \geq 1$ once $\psi_{0}$ is specified. Using the mode expansion for $\psi$ and the saddle-point approximation \cite{Campiglia:2015qka,Campiglia:2017dpg}, one finds that the leading mode $\psi_{(0)}$ is associated with particle and antiparticle oscillators, denoted respectively by $(b_{k}, b^*_{k})$ and $(d_{k}, d^*_{k})$. Specifically,
\be
\psi_{(0)}(\rho, \hat{x}) =
\begin{cases}
    \frac{\sqrt{M}}{2(2\pi)^{3/2}} e^{i\pi/4} b_{M\vec{Y}}\,,  & \quad \alpha = iM \, \\
    i\frac{\sqrt{M}}{2(2\pi)^{3/2}} e^{i\pi/4} d^*_{M\vec{Y}}\,,  & \quad \alpha = -iM \,.
\end{cases}
\ee
Furthermore, the current admits the expansion series
\be
J=\sum^{\infty}_{N=0}\frac{1}{\tau^{N+3}} J_{N}(Y)\,, \quad J_{N}(Y)=g \sum^{N}_{n=0}\psi_{(N-n)}\psi^{*}_{(n)}\,.
\ee
Explicitly, we can write down the first four currents,
\be
\label{Js}
\begin{split}
J_0(Y)&=g \,  \psi_0^*\, \psi_{0} \,,\\
J_1(Y)&=\frac{g}{2\alpha} \left[\Delta \psi_0^*\, \psi_0 -\psi_0^* \Delta \psi_0\right] \,,\\
J_2(Y)&=\frac{g}{8\alpha^2} \left[ \Delta^2 \psi_0^* \psi_0+ \psi_0^* \Delta^2 \psi_0-2(\psi_0^* \Delta \psi_0+\Delta\psi_0^*  \psi_0+\Delta\psi_0^*\Delta\psi_0 
) -3\psi_0 \psi_0^*  \right] \,,\\
J_3(Y)&=\frac{g}{48 \alpha^3} \left[\Delta^3 \psi_0^* \psi_0- \psi_0^* \Delta^3\psi_0+8(\psi_0^*\Delta^2 \psi_0-\Delta^2 \psi_0^*\psi_0) - 3 (\Delta^2 \psi^*_0 \Delta \psi_0 - \Delta \psi^*_0 \Delta^2 \psi_0  )\right]\,.
\end{split}
\ee
For the quantization of these operators, we use the normal order prescription after exchanging the conjugate $*$ by the adjoint operation $\dagger$. The corresponding non-vanishing commutation relations are
\be
\label{com}
\left[b_{\vec{p}},b^{\dagger}_{\vec{k}}\right]=\left[d_{\vec{p}},d^{\dagger}_{\vec{k}}\right]=(2p^{0})(2\pi)^3 \delta^{(3)}(\vec{p}-\vec{k})\,.
\ee

\section{Sub$^n$-Soft Theorems from Conservation laws}
\label{s4}
The conservation of the constructed charge $Q_{(m)}$ suggests that it commutes with the $S-$matrix at the quantum level. In a scattering process involving massive particles, this indicates   
\be
\label{LAeq}
\langle {\rm out}|[Q^{\rm soft}_m,S]| {\rm in}\rangle+\langle {\rm out}|[Q^{\rm hard}_m,S]| {\rm in}\rangle=0\,,
\ee
where we have defined $|{\rm in}\rangle$ and $\langle {\rm out}|$ are asymptotic states associated to a process involving $N$ massive particles. 

In what follows, we will derive each expression of the above equation using the explicit representations \eqref{qS} for the soft charge and \eqref{hardip} for the corresponding hard contribution. 
\subsection{Soft contribution}
The soft pieces at future and past null infinity are given by
\be
\begin{split}
\label{softop}
Q_{m}^{\text{soft}+}\left[f_+\right]&=\frac{-1}{2^{m}\left(m!\right)^{2}}\int_{\dI^+}d^{2}x\,du\sqrt{\gamma}\dD^{(m)}f_+ \, u^{m}\, \d_u \phi_0\,,\\
Q_{m}^{\text{soft}-}\left[f_-\right]&=\frac{(-1)^m}{2^m (m!)^2} \int_{\dI^-}d^{2}x\,dv\sqrt{\gamma}\dD^{(m)}f_- \, v^{m}\, \d_v \phi_0\,.
\end{split}
\ee
To evaluate these expressions, we need to study the role of the function $f$ and also determine the action of $\phi_0$ on the asymptotic states.

\paragraph{Choosing $f$:}  Let us first recall that the function $f(\hat{x})$ --defined on $\mathscr{I}^{\pm}$ through the antipodal map \eqref{anti}-- parametrizes an infinite family of conserved charges. At this point, we will choose this freedom to eliminate the integral on the sphere to represent the insertion of a subleading soft operator. The special choice of $f$ made below should therefore be viewed only as a convenient projection. It does not single out a preferred symmetry from the outset, nor does it discard the remaining charges. Rather, it provides a useful representation of the same angle-dependent conservation law in a basis adapted to the sub$^{n}-$soft expansion. Similar projection procedures are standard in the asymptotic symmetries literature; see, for instance, the construction of the differential operator obtained by smearing the Ward identity in Eq. (3.17) of \cite{Campiglia:2018dyi}.
To isolate this contribution from the soft part of the charge, we use the function
\be
\label{p00}
f(\hat{x},\hat{y})=  \dD^{(m)} \left[ (-2 q(x)\cdot q(y))^{m-1}\log(-2 q(x) \cdot q(y)) \right]\,,
\ee
with $q(x)=(1,\hat{x})$ a null vector. One can verify that this function satisfies (see appendix \eqref{AppG} for more details on this point), 
\be 
\label{f00} \dD^{(m)}f(\hat{x},\hat{y})=  4^{m} [(m-1)!]^2  \frac{\pi}{\sqrt{\gamma}}\dD^{(m)}\delta^{(2)}(x-y) \,.
\ee 
Before continuing, recall that when applying the operator \eqref{softop} on the incoming states, we need to take into consideration the matching condition \eqref{anti} for the parameter $f$. In particular, we use that
\be
f_{+}(\hat{x})= f(\hat{x},\hat{y})\, \quad f_{-}(\hat{x})=(-1)^m f(-\hat{x},\hat{y})\,.
\ee
Plugging this information into the soft charge yields
\begin{equation}
\begin{split}
Q_{m}^{\text{soft}+}\left[f_{+}\right]&=-\frac{2^{m} \pi}{m^{2}} \dD^{(m)}\int^{\infty}_{-\infty} du \, u^{m} \d_u \phi_0(u,\hat{y})\,,\\
Q_{m}^{\text{soft}-}\left[f_{-}\right]&=\frac{2^{m} \pi}{m^{2}}\dD^{(m)}\int^{\infty}_{-\infty} dv \, v^{m} \d_v \phi_0(v,-\hat{y})\,.
\end{split}
\end{equation}

\paragraph{Mode expansion for $\phi_0$:} 
The above operators create states of order $\omega^{m}$,  where $\omega$ is the energy of the massless scalar particle. To effectively recognize this subleading contribution in the energy, $Q_{m}^{\text{soft}}$ must be expressed in terms of oscillator mode $(a_k,a^\dagger_k)$ of $\phi_0$. They are related to the free massless field by
\be
\label{Free1}
\phi(x)=\int \frac{d^3 {\bf k}}{(2\pi)^3 2 k^0} \left(a_k e^{i k\cdot x} + a^{\dagger}_k e^{-i k\cdot x} \right)\,.
\ee
It is convenient to parameterize the momentum and the coordinate $x$ at null infinity as
\be
k^\mu=\omega (1,\hat{q})\,, \quad x^\mu=\left(u+r,r \hat{x} \right)\,. 
\ee
For large values of  $\omega r\gg1$, one applies the stationary phase approximation, finding an expansion consistent with the scalar field fall-off at null infinity. This produces
\be
\label{qq}
\phi_{0}(u,\hat{x})=\frac{1}{8\pi^2i}\int^\infty_0 d\omega\, (a(\omega \hat{x}) e^{-i\omega u} -a^\dagger(\omega \hat{x})e^{i \omega u} )\,.
\ee
Now we use general formulas involving the subleading $n-$soft operator at future and past null infinity. They can be expressed in terms of oscillators, yielding
\be
\label{a1}
\int^{\infty}_{-\infty} du\, u^{n} \d_u \phi_0(u,\hat{x})=-\frac{1}{8\pi i^{n}} \lim_{\omega\to 0} \d_{\omega}^n\left[ \omega(a_{\rm out}(\omega \hat{x}) + (-1)^{n}a_{\rm out}^\dagger(\omega \hat{x}))\right] \,, 
\ee
and analogously, one finds that at past null infinity 
\be
\label{a1.5}
\int^{\infty}_{-\infty} dv\, v^{n} \d_v \phi_0(v,\hat{x})=\frac{1}{8\pi i^n} \lim_{\omega\to 0} \d_{\omega}^n\left[ \omega(a_{\rm in}(-\omega \hat{x}) + (-1)^{n}a_{\rm in}^\dagger(-\omega \hat{x}))\right]\,.
\ee
Using all these relations and applying crossing symmetry, we can express the action of the subleading soft operator in a scattering process as 
\be
\label{QS2}
\langle {\rm out}|[Q^{\rm soft}_m,S]| {\rm in}\rangle= \frac{2^{m} }{4 i^m m^{2}}  \dD^{(m)}\lim_{\omega \to 0} \d^{m}_{\omega}\left(\omega\,  \langle  {\rm out}|  \,a^{\rm out}(\omega \hat{y}) S| {\rm in}\rangle \right) \,.
\ee

\subsection{Hard contribution}
Continuing with the evaluation of \eqref{LAeq}, we now consider the hard contribution. Using the explicit form of the hard charge \eqref{hardip}, one finds
\be
\langle {\rm out}|[Q^{\rm hard}_m,S]| {\rm in}\rangle=\int d^3 V f_{m}(Y;q) \langle {\rm out}|[J_m,S]| {\rm in}\rangle\,.
\ee
This expression again depends on two variables: the smeared function on the sphere, which is now projected onto the hyperboloid $f_m(Y;q)$, and the action of the higher mode currents $J_m(Y)$ on the external states.

The smearing function at future time-like infinity $f^+_{m}(Y;q)$ is obtained after inserting the parameter $f$ \eqref{p00} in the definition \eqref{hardip2}. The details of this computation have been worked out in the appendix \eqref{hyp} using Feynman parameters. The result is
\be
\label{lamg}
f^{+}_{m}(Y;q)=\frac{1}{2m} \dD^{(m)}\left[ (-2q\cdot Y)^{m-1} \log(-2q\cdot Y)\right]\,.
\ee
There is a similar expression for this function at past time-like infinity that will be presented at the end of \eqref{424}. It is worth noticing that the presence of the pole $q\cdot Y=0$ will be central in the derivation observed in the diagrammatic result \eqref{subsfeyn}.

Applying the operators $J_m$ to physical states requires a detailed understanding of the Fock space from which they are constructed. In Section \eqref{sub-mass}, we explicitly constructed these currents for the theory under consideration in terms of the free data $\psi_{(0)}$. 

To better understand its action, we will diagonalize them for $m = 1,2,3$ on single-particle states. We focus on the operators associated with particles $b_k, b^{\dagger}_k$ for concreteness. The extension to anti-particles $d_k, d^{\dagger}_k$ is straightforward. In the last subsection, we provide a general argument to  obtain the action of the hard charge for any value of the integer $m$.

\subsubsection{First subleading hard charge}
The first current can be written as a divergence
\be
J_1(Y)=-i \frac{g}{8 (2\pi)^3} \nabla^{a}\left[\nabla_a b^{\dagger}_{M\vec{Y}}\, b_{M\vec{Y}}- b^{\dagger}_{M\vec{Y}} \,\nabla_a b_{M\vec{Y}}\right]\,.
\ee
We can then show that after integrating by parts, using the commutation relations \eqref{com}   and $\Delta f_{1}=0$, we find
\be
Q^{\rm hard}_1|k\rangle=-i\frac{g}{2M} \nabla_a f_{(1)}(k/M;q) \, \nabla^a |k\rangle\,,
\ee
where the covariant derivative $\nabla_a$ is defined on the hyperboloid. Interestingly, the above operation acts as a diffeomorphism on a scalar with parameter $f_{(1)}$. We can further develop this result after using \eqref{lamg}, leading to  
\be
Q^{\rm hard}_1| k\rangle= -i \frac{g}{4 M^2 }  D^2 \left(\frac{q\cdot \nabla_a k }{q\cdot k} \, \right)\nabla^a k^{\mu} \frac{\d}{\d k^{\mu}} | k\rangle\,.
\ee
Finally, using the identities \eqref{aaYY}, we find that $(q\cdot \nabla_a k) \nabla^a k^{\mu}=M^{2}q^{\mu}+(q \cdot k) k^{\mu}$, yields
\be
Q^{\rm hard}_1| k\rangle= -i \frac{g }{4} D^2\left(  \frac{q^{\mu} }{q\cdot k}   \frac{\d}{\d k^{\mu}}\right)  | k\rangle\,.
\ee

\subsubsection{Second subleading hard charge}
We continue computing the second subleading charge.
We will first express the external current in terms of the oscillator operators. Using the expression in \eqref{Js} for $J_2(Y)$ and that $\Delta f_{2}=3 f_{2}$, we can further simplify the hard charge by considering the identity obtained through integration by parts
\begin{multline}
\int d^3V \,f_{2} \left[ \Delta^2 b^{\dagger}_{M\vec{Y}}\, b_{M\vec{Y}}+ b^{\dagger}_{M\vec{Y}}\, \Delta^2 b_{M\vec{Y}} -3 b^{\dagger}_{M\vec{Y}}\, b_{M\vec{Y}}  \right]\\
= 2\int d^3V \, \left[ f_{2}\left(\Delta b^{\dagger}_{M\vec{Y}}\, b_{M\vec{Y}}+ b^{\dagger}_{M\vec{Y}}\, \Delta b_{M\vec{Y}}+ \Delta b^{\dagger}_{M\vec{Y}} \Delta b_{M\vec{Y}} - \nabla_a b^{\dagger}_{M\vec{Y}}\, \nabla^a b_{M\vec{Y}}\right)\right. \\  
\left.  +\nabla_a f_{2} \left(\Delta b^{\dagger}_{M\vec{Y}} \nabla^a b_{M\vec{Y}}+
\nabla^a b^{\dagger}_{M\vec{Y}}  \Delta  b_{M\vec{Y}}
 \right) \right]\,.
\end{multline}
The above expression rewrites the bi-Laplacian operator in terms of Laplacians and covariant derivatives, providing a decomposition that facilitates the subsequent analysis. In fact, using the above, we find that the normal-ordered hard charge can be written as
\be
Q^{\rm hard}_2=-\frac{g}{16 M (2\pi)^3} \int d^3V \, \left[ \nabla_a f_{2} \left(\Delta b^{\dagger}_{M\vec{Y}} \nabla^a b_{M\vec{Y}}+ \nabla^a b^{\dagger}_{M\vec{Y}} \Delta b_{M\vec{Y}}  \ \right) -f_{2} \nabla_a b^{\dagger}_{M\vec{Y}} \nabla^a b_{M\vec{Y}} \right]\,.
\ee
We now examine its action on a single-particle state of momentum $k^{\mu}$. The resulting expression simplifies to
\be
Q^{\rm hard}_{2} |k \rangle= \frac{g}{4 M^3}  \left( f_{2} \Delta- \nabla_a \nabla_b f_{2} \nabla^a \nabla^b \right)|k\rangle \,,
\ee
which provides the hard action of the charge through second-order differential operators defined on the hyperboloid. We now use 
\be
f^{(2)}(Y;q)=\tfrac{1}{4}\dD^{(2)}\left[ (-2q\cdot Y)\log(-2q\cdot Y) \right]\,,
\ee
where the operator on the sphere acts only on the vector $q(x)$. 
Explicitly computing the second derivatives, one finds:
\be
\begin{split}
\nabla_a \nabla_b f_{2}&= -\tfrac{1}{2}\dD^{(2)} \left[ \frac{(q\cdot \nabla_a Y)(q\cdot \nabla_b Y)}{q\cdot Y} \right] +g_{ab} f_{2}\,,\\
\nabla^a \nabla^b |k\rangle &=g^{ab} k^\mu \frac{\d}{\d k^{\mu}} |k\rangle + \nabla^a k^\mu \nabla^b k^\rho \frac{\d^2}{\d k^\mu \d k^\rho}|k\rangle  \,.
\end{split}
\ee
The final expression can be further reduced, and one finds
\be
Q^{\rm hard}_2|k\rangle =\frac{g}{8} \dD^{(2)} \left[ \frac{1}{q\cdot k} \left( q\cdot \frac{\d}{\d k}\right)^2 \right]|k\rangle  \,.
\ee

\subsubsection{Third subleading hard charge}
The operator for this case can be expressed as
\be
Q^{\rm hard}_3=\frac{g}{24\alpha^3}\int d^3 V  \left[\nabla_a f_3 \left( \Delta^2 \psi_0^{\dagger} \nabla^a \psi_0 -\nabla^a \psi_0^{\dagger}  \Delta^2 \psi_0 \right)  - f_3(\Delta^2 \psi^\dagger_0 \Delta \psi_0 - \Delta \psi^\dagger_0 \Delta^2 \psi_0  ) 
\right]\,,
\ee
where we have performed integrations by parts to replace the cubic Laplacian $\Delta^{3}$ in terms of lower-order differential operators on the hyperboloid. Here, the function $f_3$ now satisfies $\Delta f_3= 8 f_3$.

Following the same steps as before, and after a long albeit straightforward computation, one finds a third derivative action given by
\be
Q^{\rm hard}_3|k\rangle =-\frac{ig}{12 M^{4}}\left[\nabla_c\nabla_b \nabla_a f_3 \nabla^c\nabla^b\nabla^a  + 4 \nabla_a f_3 \nabla^a 
- 4 \nabla_a  f_3  \nabla^a \Delta \right]|k\rangle\,,
\ee
and using the explicit form of $f_3$, we get
\be
Q^{\rm hard}_3|k\rangle =\frac{ig}{9}\dD^{(3)} \left[\frac{1}{q\cdot k} \left(q\cdot \frac{\d}{\d k}\right)^3 \right]|k\rangle\,.
\ee

\subsubsection{The general case}
\label{424}
We now provide a general argument to argue that the action of the hard charge takes the following form for an arbitrary value of $m$
\be
Q^{\rm hard}_m|k\rangle=-\frac{(2i)^{m}}{8 m^2}    \frac{ g}{q\cdot k} \left( q\cdot \frac{\d}{\d k}\right)^m |k\rangle\,. \label{ref}
\ee
To do so, we will analyze the scaling properties of the $Q^{\rm hard}_n$ under the transformations $q \to \lambda q$. We have previously shown that the $n-$th subleading hard charge is given by
\be
\label{qhardd}
Q^{\rm hard}_n[q]=  \frac{1}{2n} \dD^{(n)} \int d^3V  \left[ (-2q\cdot Y)^{n-1} \log(-2q\cdot Y)\right] J_n(Y)\,.
\ee
We want to determine how this operator acts on the states. Notice that 
\be
\label{qhard}
Q^{\rm hard}_n[\lambda q]=\lambda^{n-1} Q^{\rm hard}_n[q]\,,
\ee
where we have used that $\dD^{(n)}\left[(-2q\cdot Y)^{n-1} \right]=0$. The above equation implies that the action of the hard charge is a homogeneous function of degree $n-1$ in the vector $q$. It follows from Lorentz invariance that the functional dependence can only be expressed in terms of the combinations $q \cdot k$ and $q \cdot \d_k $. Therefore, the most general form of this operator, consistent with equation \eqref{qhard} and having at least a single pole, is
\be
Q^{\rm hard}_n = 
\cF\left(\frac{q\cdot \d_k}{q\cdot k }\right)(q\cdot \d_k)^{n-1} \,,
\ee
where $\cF$ represents an arbitrary function. A comparison with the previously established hard actions for cases $n=1$, $n=2$, and $n=3$ indicates that the arbitrary function can only exhibit linear dependence, $\cF(x)=c_n x$, where $c_n$ is a constant that cannot be determined through this analysis. However, looking at the sequence provided by $c_1=-g\tfrac{i}{4}$, $c_2=g\frac{1}{8}$, $c_3=g\tfrac{i}{9}$, we can infer that 
\be
c_n=-g\frac{(2i)^{n}}{8 n^2}\,.
\ee

\subsubsection*{Hard charge in the scattering process}
Finally, we can extend the result to the full scattering process, with $N_1$ incoming particles and $N_2$ outgoing particles. We parameterize outgoing states using the time-like vector \eqref{Ydef}, $k^{\rm out}_j=M Y(\rho_j,\hat{n}_j)$. In this case, we must act with states on the left, which adds an extra factor $(-1)^{m}$. This yields
\be
\label{QhOUT}
\langle {\rm out}|Q^{\rm hard}_m= -g\frac{2^{m}(-i)^m}{8 m^2}  \dD^{(m)}\sum^{N_1}_{j=1} \frac{1}{q\cdot k^{\rm out}_j} \left( q\cdot \frac{\d}{\d k^{\rm out}_j}\right)^m  \langle  {\rm out}| \,.
\ee
For the incoming states, we have to use the matching condition \eqref{anti} that produces the following smearing function on $i^{-}$ 
\be
f^{-}_m(X;q)=\frac{(-1)^{m}}{2m} \dD^{(m)}\left[ (-2q(-\hat{x})\cdot X)^{m-1} \log(-2q(-\hat{x})\cdot X)\right],
\ee
where $X(\rho,\hat{n})=-(\sqrt{1+\rho^2},-\rho \hat{n})$ corresponds to the normal vector pointing towards $i^{-}$ antipodally related with the corresponding vector field $Y$ defined at $i^{+}$. By using that $$X(\rho,\hat{n})\cdot q(-\hat{x})=-Y(\rho,\hat{n})\cdot{q}(\hat{x})\,,$$ we can infer that the effect of the hard action on the ingoing states with momentum $k^{\rm in}_j= M Y(\rho_j,\hat{n}_j)$ gives
\be
\label{QhIN}
Q^{\rm hard}_m| {\rm in}\rangle=-(-1)^m g \frac{(2i)^{m}}{8 m^2}  \dD^{(m)}\sum^{N_2}_{j=1} \frac{(-1)^{m+1}}{q\cdot k^{\rm in}_j} \left( q\cdot \frac{\d}{\d k^{\rm in}_j}\right)^m | {\rm in}\rangle  \,.
\ee
Therefore, using the total number of particles $N=N_1+N_2$, 
\be
\label{QSH}
\langle {\rm out}|[Q^{\rm hard}_m,S]| {\rm in}\rangle= -g\frac{2^{m}}{8 m^2 i^{m}}  \dD^{(m)}\sum^{N}_{j=1} \frac{ \eta_j^{m+1}}{q\cdot k_j} \left(q\cdot \frac{\d}{\d k_j}\right)^m  \langle  {\rm out}|S| {\rm in}\rangle  \,,
\ee
where $k_j=k^{\rm in}_j$ corresponds to $\eta_j=-1$ and  $k_j=k^{\rm out}_j$ for $\eta_j=1$.  Once the above equation and its soft contribution \eqref{QS2} are replaced back in \eqref{LAeq}, we precisely find the diagrammatic result \eqref{subsfeyn}.

\section{Discussion}
\label{s5}
We have developed a systematic method for constructing an infinite set of conserved quantities in scattering processes that involve the emission or absorption of a massless scalar particle. We have proven that the Ward identities associated with conserved charges give rise to an infinite set of subleading soft relations, which determine the ``factorizable''  part of the amplitude for the emission or absorption of a single massless scalar. In other words, this new infinite tower of conserved charges determines all the subleading soft theorems in the series expansion with respect to the frequency of the massless particle. Crucially, the charges remain finite at every step, and as a result, no holographic renormalization is required.

In particular, our analysis was based on the study of a theory in which a massless scalar field is coupled to a massive scalar field through a scalar version of Yukawa theory. However, one can expect that these results will naturally extend to other types of models describing hard particles. For example, one could study the changes in our derivation when considering a massless scalar theory coupled to a massive field of any spin via a Yukawa-like interaction
\be
\cL_{\rm int}=g \phi \Psi_{\mu_1 \mu_2\cdots \mu_s}\Psi^{\mu_1 \mu_2\cdots \mu_s}\,,
\ee
where $\Psi_{\mu_1 \mu_2\cdots \mu_s}$ is a completely symmetric tensor with the appropriate transversality and trace condition.

We did not explicitly compute the algebra of this new infinite set of charges. However, there are good arguments to believe that it must be abelian, possibly with a central extension. This follows from the fact that the commutator of free massless scalar fields at unequal times is given by the Pauli-Jordan function, which is a $c-$number. Since the charges are constructed from different orders of the scalar field in the radial expansion, their commutator can be at most a $c-$number, thereby defining a central extension. In the future, it would be interesting to explicitly compute this central charge, if it exists.

In our analysis, we restricted the asymptotic treatment to the branch that excludes radial logarithmic behavior
\be
\phi=O\left(\frac{\log(r)}{r}\right),
\ee
even though such solutions are admitted \cite{Fuentealba:2024lll}. 

It would be interesting to explore the role of logarithmic branches in the context of soft theorems and, in particular, whether they are connected in some way to the logarithms that appear in the classical formulation of soft theorems (see, e.g., \cite{Sahoo:2018lxl,Sahoo:2020ryf,Sen:2024bax,Karan:2025ndk,Duary:2025siq}).

So far, our computations have been carried out at tree level. However, it is worth exploring the inclusion of loop corrections\footnote{ This analysis would require a model that ensures that the masslessness of the scalar field is preserved under renormalization.}. In this context, it would be interesting to investigate the connection between such corrections and possible \(\log(\omega)\) terms in the amplitudes, as well as their relation to trajectory tails. Specifically, in  \cite{Duary:2025siq}, it has been shown that the logarithmic dependence on the retarded time shows up in the waveform of a massless scalar due to its interaction with classical point particles. One would therefore expect that relaxing condition (3.10) in our construction would naturally allow such terms to appear in the derivation of the charges, altering their structure compared to the tree-level case.

It is important to emphasize that studying scalar fields, at least in their standard formulation, has the advantage of not involving gauge symmetries. This makes them valuable toy models for gaining deeper insights into the infrared structure of theories with massless particles, as well as the role of the tower  of conserved charges associated with subleading soft theorems. This approach, therefore, offers a different perspective on the problem compared to electromagnetism and gravitation. For example, the tower of conserved charges does not appear to correspond to symmetries of the classical Lagrangian. In this regard, it is natural to ask whether the infinite conserved charges are related to Noether symmetries in some yet unknown way. In \cite{Campiglia:2018see,Francia:2018jtb}, the charge associated with the leading soft theorem was described using a dual two-form gauge potential for the scalar field. Generalized asymptotic symmetries for $D-2$ forms in $D-$dimensions have been studied in recent work \cite{Romoli:2024hlc,Manzoni:2025wyr}. 
 In our setup, the scalar field obeys the free wave equation near spatial infinity $i^0$, and the massive sources do not reach this region. Therefore, the relevant asymptotic dynamics coincides with that of a free two-form field in $D=4$. After symplectic renormalization, the charges constructed in \cite{Romoli:2024hlc} agree with those found here when the form field is represented in terms of its scalar dual. It would be interesting to further explore how our subleading charges can be systematically derived from finite asymptotic symmetries of the two-form gauge field.

It may be useful to give an interpretation to the fact that, according to Eq.~\eqref{Qms2}, the $n$-th conserved charge depends linearly on the $n$-th order term in the expansion of the scalar field in powers of $1/r$. When analyzing the factorizable part of the amplitude for the emission of massless particles with higher energies, it becomes necessary to consider charges associated with field components located deeper in the bulk. A heuristic explanation for this effect is that the notion of a ``large $r$ expansion'' should be understood in relation to a characteristic length scale, in this case, the wavelength of the emitted massless particle. In other words, the relevant dimensionless parameter is $\omega r$. For instance, for a fixed value of this parameter, large $r$ corresponds to small frequencies, which is the regime typically considered in leading soft theorems. Consequently, as the energy of the emitted particle increases, a deeper region of the bulk must be examined to describe the process accurately. This suggests that, following some of the ideas outlined in \cite{Conde:2016rom}, the infinite tower of charges in this theory—as well as in electromagnetism and gravity—could be reinterpreted as a certain surface integral defined at a finite radius  $r_b$, such that the leading soft theorems are recovered in the limit when this ``boundary'' approaches infinity, $r_{b}\rightarrow{\infty}$. 

 Finally,  it is natural to ask how the infinite tower of conserved charges constructed in this work is realized in the celestial framework. As argued after Eq. \eqref{deltaY}, the sub$^n$-leading soft scalar operator has conformal weights $(\tfrac{n+1}{2},\tfrac{n+1}{2})$ and thus the corresponding Ward identity does not take the standard form of a holomorphic current on the celestial sphere. This suggests that the symmetry interpretation of scalar soft theorems in two dimensions is more subtle than in gauge or gravitational theories. Our construction, however,  follows from the conservation of surface charges derived from the asymptotic expansion and matching conditions at spatial infinity. From this perspective, it would be interesting to investigate whether an appropriate celestial realization of these charges may involve descendant operators or smeared insertions on the sphere, rather than the primary sub$^{n}$-soft operators themselves. Understanding how the Hamiltonian matching at $i^0$ translates into celestial correlators may provide further insight into the infrared structure of scalar theories and their relation to generalized asymptotic symmetries.\footnote{We thank the anonymous referee for pointing out this interesting point.}  We expect to address some of these aspects in future work.

\vspace{1cm}
\paragraph{Acknowledgements}
~\\[8pt]
It is a pleasure to thank Glenn Barnich,  Miguel Campiglia, Laurent Freidel, Marc Henneaux, Yorgo Pano, Francisco Rojas, Jakob Salzer, and Ali Seraj for their insightful comments and fruitful discussions.We also thank the anonymous referee for pointing out the contribution of the
$l<k$ branch to the coefficient in \eqref{changed}. The work of H.G. and M.B. is funded by the ANID FONDECYT grant 1230853.  The research of A.P. is supported by the ANID FONDECYT grant 1260427.

\appendix

\section{Green's functions on the sphere}
\label{AppG}
We consider the following differential operator defined on the sphere for $n>0$
\be
\dD^{(n)} \equiv \prod^{n}_{k=1}(D^2 +k(k-1))\,,
\ee
where $D^2$ is the Laplacian on the sphere. Since this operator annihilates the spherical harmonics with $l < n $, the corresponding Green's equation is defined by
\be
\label{eq:GA}
\dD^{(n)} \omega_n(x,x')= \delta_{(n)}(x-x') \,,
\ee
where the function $\delta_{(n)}$ is related to the Dirac delta on the sphere $\delta^{(2)}$ by subtracting the harmonics with $l<n$ from the following completeness relation
\be
\label{eq:A3}
\frac{1}{\sqrt{\gamma}}\delta^{(2)}(x-x')=\sum^{\infty}_{l=0} \sum^{m=l}_{m=-l} Y_{lm}(\hat{x})Y^{*}_{lm}(\hat{x}')\,.
\ee
Doing so and applying the addition theorem\footnote{$P_l(\hat{x}\cdot\hat{y})=\tfrac{4\pi}{2l+1}\sum^{l}_{m=-l} Y_{lm}(\hat{x})Y^{*}_{lm}(\hat{y})$.}, the explicit expression for this function is
\be
\label{eq:A4}
\delta_{(n)}(x-x')=\frac{1}{\sqrt{\gamma}}\delta^{(2)}(x-x')-\frac{1}{4\pi}\sum^{n-1}_{l=0} (2l+1) P_{l}(\hat{x}\cdot\hat{x}')\,,
\ee
where $P_{(l)}(x)$ are the Legendre polynomials of degree $l$. Using the normalization established above, the solution to \eqref{eq:GA} is given by
\be
\omega_n(x,x')=\alpha_n (-2 q\cdot q')^{n-1}\log(-2 q \cdot q')\,,\quad \alpha_n=\frac{1}{ 4^{n} ((n-1)!)^2 \pi}\,,
\ee
with $q\equiv(1,\hat{x})$ and $q'\equiv(1,\hat{x}')$ two null vectors.

By construction, the function containing the harmonics $l< n$  
\be
C_{n}(x,x'):=\frac{1}{4\pi}\sum^{n-1}_{l=0} (2l+1) P_{l}(\hat{x}\cdot\hat{x}')\,,
\ee
satisfies 
\be
\label{newB}
\dD^{(n)} C_n=0\,.
\ee
Therefore, a representative of Green's functions on the sphere is defined by
\be
f_n(x,x'):=\dD^{(n)} \omega_n(x,x')\,,
\ee
which does not contain the first $n$ harmonics and therefore 
\be
\label{eq:GA3}
\dD^{(n)}\left( f_{n}(x,x')- \frac{1}{\sqrt{\gamma}}\delta^{(2)}(x-x')\right)=0 \,.
\ee

\subsection{Operators on the sphere}
\label{ops2}
The normal vector $\hat{q}$ on the sphere satisfies some important properties. Its covariant derivative acts as a projector, therefore
\begin{align}
    \gamma^{AB}D_A \hat{q}^{i} D_B \hat{q}^{j}=\delta^{ij}-\hat{q}^{i}\hat{q}^{j}\, \quad D_A D_B \hat{q}=-\gamma_{AB} \hat{q}\,.
\end{align}
In order to demonstrate the statement \eqref{eqYlm}, we can first prove, using the previously established relations, that 
\begin{align}
    (D^2 + 6) (\hat{q}^{i}\hat{q}^{j})=\delta^{ij}.\label{basecase}
\end{align}
Note that the operator corresponds to $l=2$ harmonics. Thus, it follows that when you apply an operation to a polynomial of degree $l$, the resulting expression will be of a lower degree than $l$. To demonstrate this, we will use induction, starting with the assumption
\begin{align}
    (D^2+ l(l+1))  (\hat{q}^{i_1}\cdots\hat{q}^{i_l})= P^{-}_l(q), \label{hypo}
\end{align}
with $P^{-}_l(q)$ representing a polynomial of $q$ of order lower than $l$. 

The initial case was already computed in \eqref{basecase}. Following the induction procedure, assuming that it holds for $l$, we need to prove that it implies the same for $l+1$. 
Specifically, we have to assume \eqref{hypo} and then prove
\begin{align}
 (\cD^2 + (l+1)(l+2))(\hat{q}^{i_1}\cdots \hat{q}^{i_{l+1}})= P^{-}_{l+1}(q).
\end{align}
Computing the first term then
\begin{align*}
    D^2 (\hat{q}^{i_1}\cdots \hat{q}^{i_{l+1}})=&\hat{q}^{i_{l+1}}D^2(\hat{q}^{i_1}\cdots\hat{q}^{i_l}) + 2 D_A \hat{q}^{i_{l+1}} D^A(\hat{q}^{i_1}\cdots\hat{q}^{i_{l}})+\hat{q}^{i_1}\cdots\hat{q}^{i_{l}} D^2 \hat{q}^{i_{l+1}}\,,\\
    =&-l(l+1)\hat{q}^{i_1} \cdots \hat{q}^{i_{l+1}} + \hat{q}^{i_{l+1}} P^{-}_l(q)-2 \,l\, \hat{q}^{i_1}\cdots\hat{q}^{i_{l+1}} -2 P^{-}_{l-1}(q) - 2 \hat{q}^{l}\cdots\hat{q}^{i_{l+1}}\\
    =&-(l+1)(l+2)\hat{q}^{i_1}\cdots \hat{q}^{i_{l+1}}+P^{-}_{l+1}(q),
\end{align*}
where on the second line we used that $P^{-}_{l-1}(q)$ is another polynomial in $q$ of order $l-1$. So, putting this form back into the hypothesis, we obtain the expected result.

Now, using \eqref{hypo}, the application of the operators $D^2+j(j+1)$ with $j=l-1,\cdots,0$  lowers the degree of the polynomial until it vanishes, finding
\begin{align}
    (D^2 + l(l+1))(D^2 + (l-1)l)\cdots(D^2 + 2)D^2 (q^{i_1}\cdots q^{i_l})=0\,.
\end{align}

\section{$\tilde{P}$-branch and parity of the $u$-independent coefficient}
\label{app:polybranch_parity}
In this appendix, we show that the contribution of the  $l<k$-branch to the coefficient
$\phi^{(k,0)}$ obeys the same antipodal matching condition as the $l\ge k$ branch. Consider the contribution of the polynomial branch in \eqref{expansionphi}. The corresponding term in the sum near $\dI^+_-$ can be written as
\be
\frac{1}{r(-2u)^k}\left(1+\frac{u}{2r}\right)^{-k}
\tilde P^{\left(k+\frac12\right)}_{l-k}\!\left(1+\frac{u}{r}\right)
\ee
The term of our interest is the one of order $u^0$ in the large $r$ expansion. Thus, the relevant term reads
\begin{align}
    \frac{1}{r^{k+1}}\frac{(-1)^k}{2^k} \sum_{n=0}^\infty \frac{(-1)^n}{2^n} \binom{k-1+n}{k-1}\frac{1}{(k-n)!} \tilde{P}^{(k+\frac{1}{2})[k-n]}_{l-k}(1)
\end{align}
where $P^{(k) [n]}_m$ is the $n-$th derivative of $P^{(k)}_m$.  As used in the main text, we define $c_l^{(k)}$ as
\be
c_l^{(k)}
:=
\frac{(-1)^k}{2^k} \sum_{n=0}^\infty \frac{(-1)^n}{2^n} \binom{k-1+n}{k-1}\frac{1}{(k-n)!} \tilde{P}^{(k+\frac{1}{2})[k-n]}_{l-k}(1).
\ee
Similarly, near $\dI^-_+$ the corresponding term of the sum goes as
\be
\frac{1}{r(2v)^k}\left(1-\frac{v}{2r}\right)^{-k}
\tilde P^{\left(k+\frac12\right)}_{l-k}\!\left(-1+\frac{v}{r}\right)\,,
\ee
Expanding for large $r$ and isolating the term that goes as $v^0$ yields
\be
\frac{1}{r^{k+1}}\frac{1}{2^k} \sum_{n=0}^\infty \frac{1}{2^n} \binom{k-1+n}{k-1}\frac{1}{(k-n)!} \tilde{P}^{(k+\frac{1}{2})[k-n]}_{l-k}(-1)\,.
\ee
Analogously to $c_l^{(k)}$, we define
\be
 d^{(k)}_l:=\frac{1}{2^k} \sum_{n=0}^\infty \frac{1}{2^n} \binom{k-1+n}{k-1}\frac{1}{(k-n)!} \tilde{P}^{(k+\frac{1}{2})[k-n]}_{l-k}(-1)\,.
\ee
Now, using that the polynomial branch $\tilde{P}$ has definite parity,
$\tilde{P}^{(\lambda)}_{n}(-s)=(-1)^{n}\,\tilde{P}^{(\lambda)}_{n}(s)$ with
$\lambda=k+\tfrac{1}{2}$ and $n=l-k$\footnote{See Eq.~(A.19) of~\cite{Henneaux:2018mgn}; see also~\cite{Fuentealba:2024lll}.}, one has 
\be
\tilde P^{\left(k+\frac12\right) [k-n]}_{l-k}(1)
=
(-1)^{l-n}
\tilde P^{\left(k+\frac12\right)[k-n]}_{l-k}(-1)\,,
\ee
which proves the relation used in the main text for the parity condition 
\begin{align}
    d^{(k)}_l=(-1)^{l-k}\;c^{(k)}_l\,.
\end{align}

\section{Green's functions and differential operators}\label{GTN}
\subsection{More on the hyperbolic embedding}
We parameterize the spacetime close to time-like infinity through hyperboloids with a normal vector $Y$ satisfying 
\be
\label{aaYY}
Y^2=-1\,, \quad  \Delta Y^{\mu}=3Y^{\mu}\,, \quad \nabla_a Y_\mu \nabla^a Y_\nu=\eta_{\mu \nu} +Y_\mu Y_\nu\,, \quad \nabla_a \nabla_b Y^\mu=g_{ab} Y^\mu\,.
\ee
In $y^a$ coordinates, defined after \eqref{GH}, the volume element reads 
\be
d^4 x= \tau^3 \sqrt{\gamma} \frac{\rho^2}{\sqrt{1+\rho^2}}d \tau  d\rho d^2 x \equiv \tau^3 d \tau d^3 V\,.
\ee
Using the $Y$-parameterization, we find that integration over the hyperboloid is given by the measure
\be
\int d^3 V = \int \frac{d^3 Y}{Y^0}.
\ee
We can then infer that the invariant Dirac delta on the hyperboloid is 
\be
\delta_{\cH}\equiv Y^{0} \delta^{(3)}(\vec{Y})\,.
\ee
Some useful relations valid for any function $\Omega$ are
\be
[\Delta,\nabla^a]\Omega=-2\nabla^a \Omega\,,\quad [\Delta,\nabla^a \nabla^b]\Omega=-6\nabla^a \nabla^b\Omega+2 g_{ab}\Delta \Omega\,.
\ee

\subsection{Null to time-like infinity Green's functions}

\label{hyp}
To obtain the null to time-like infinity propagator, it is helpful to consider the integral,
\be
I_n[q,Y]=\frac{1}{2\pi}\int d^2 x' \omega_n\left(x,x'\right) G_{n}(Y,q')\,.
\ee
Using the identity
\be
\log(x)=-\lim_{h\to0} \frac{x^{-h}-1}{h}\,,
\ee
we find
\be
\label{inqy}
\begin{split}
I_n[q,Y]&=-\frac{\alpha_n}{2\pi}\lim_{h\to0}\int d^2 x' \sqrt{\gamma'}\frac{1}{h} \left[\frac{1}{(-2 q \cdot q')^{h-n+1}(-2q'\cdot Y)^{n+1}} - \frac{1}{(-2 q \cdot q')^{1-n}(-2q'\cdot Y)^{n+1}}\right]\,,\\
&=-\frac{\alpha_n}{2\pi}\lim_{h\to0}\frac{1}{h} \left(\cS[h-n+1,n+1]-\cS[1-n,n+1]\right)\,,
\end{split}
\ee
with
\be
\cS[a,b]=\int d^2 x' \sqrt{\gamma'}\frac{1}{(-2 q \cdot q')^{a}(-2q'\cdot Y)^{b}}\,,
\ee
that can be evaluated using one Feynman parameter $\alpha$:
\be
\frac{1}{A_1^a A_2^b}=\frac{1}{B(a,b)}\int^\infty_0 d\alpha \,  \frac{\alpha^{b-1}}{(A_1+\alpha A_2)^{a+b}}\,,\quad  B(a,b)=\frac{\Gamma(a)\Gamma(b)}{\Gamma(a+b)}=\int^{\infty}_0 dt \frac{t^{a-1}}{(1+t)^{a+b}}\,.
\ee
The latter expression yields 
\be
\cS[a,b]=\frac{4\pi}{2^{a+b}}\frac{B(\tfrac{b-a}{2},a)}{B(b,a)}\frac{1}{(-2Y\cdot q)^{a}}\,.
\ee
Plugging this result back in \eqref{inqy} and taking the $h\to0$ limit, we find
\be
I_{n}[Y,q]=\frac{\alpha_n}{2n} (-2q\cdot Y)^{n-1} \left[\log(-2q\cdot Y)+a_n\right]\,, \quad a_n=\log(2)+\tfrac{1}{2}\sum^{n-1}_{k=1}\frac{1}{k}-1\,.
\ee
Using \eqref{aaYY}, one finds that the above function fulfills the equation
\be
\Delta_H I_{n}[Y,q]=(n^2-1) I_{n}[Y,q] +2\pi \alpha_n (-2Y\cdot q)^{n-1}\,.
\ee
It is then possible to define
\be
\label{aaf}
f^{(n)}_{H}(Y)\equiv \dD^{(n)} \, I_{n}[Y,q]\,,
\ee
that  satisfies relation \eqref{a5} due to equation \eqref{newB}.

\providecommand{\href}[2]{#2}\begingroup\raggedright\endgroup


\end{document}